# Post-Experiment Forensics and Human-in-the-Loop Interventions in Explainable Autonomous Scanning Probe Microscopy

Yongtao Liu,[1,a] Maxim Ziatdinov,[1,2] Rama Vasudevan,[1] and Sergei V. Kalinin[3,b]

[1] Center for Nanophase Materials Sciences, Oak Ridge National Laboratory, Oak Ridge, TN 37830, USA

[2] Computational Sciences and Engineering Division, Oak Ridge National Laboratory, Oak Ridge, TN 37830, USA

[3] Department of Materials Science and Engineering, University of Tennessee, Knoxville, TN, 37996, USA

The broad adoption of machine learning (ML)-based automated and autonomous experiments (AE) in physical characterization and synthesis requires development of strategies for understanding and intervention in the experimental workflow. Here, we introduce and realize strategies for post-acquisition forensic analysis applied to the deep kernel learning based AE scanning probe microscopy. This approach yields real-time and post-acquisition indicators of the progression of an active learning process interacting with an experimental system. We further illustrate that this approach can be extended towards human-in-the-loop autonomous experiments, where human operators make high-level decisions at high latencies setting the policies for AE, and the ML algorithm performs low-level fast decisions. The proposed approach is universal and can be extended to other physical and chemical imaging techniques and applications such as combinatorial library analysis. The full forensic analysis notebook is publicly available on GitHub at https://github.com/yongtaoliu/Forensics-DKL-BEPS.

[a] liuy3@ornl.gov
[b] sergei2@utk.edu

Over the last several years, the attention of the scientific community has firmly been riveted to the introduction and optimization of automated experiments in areas including material synthesis, physical characterization, and microscopy. For materials synthesis, multiple approaches including pipetting robots,[1,2] self-driving labs,[3-5] and high throughput synthesis workflows have been proposed.[6-11] For characterization, several groups are now developing automated experiment approaches in areas including scanning transmission electron microscopy,[12-14] scanning probe microscopy,[15-19] neutron[20,21] and X-ray scattering.[22]

The central concept in automated experiments is the workflow,[23] defined as the sequence of steps and operations performed by the automated laboratory or the measurement tool. Generally, the workflow can combine the steps performed by human and non-human agents. For example, for many microscopes operations such as tuning the microscope are performed automatically, but the selection of regions for specific measurements are currently performed by human operators. Similarly, in materials synthesis the optimization of the specific synthesis conditions can be performed via an automated synthesis platform, but the selection of the endmembers defining the phase diagram is often manual.

Until now, the vast majority of these automated efforts have been based on human-designed workflows, with the role of any machine learning (ML) algorithm limited to (often greedy) optimization of the consecutive process steps. In this process, each step of the workflow is optimized individually, whereas possible correlations between individual steps are ignored. The recent advances in approaches such as deep kernel active learning are enabling the development of single-step beyond-human workflows, as exemplified by the property discovery in scanning probe[16,24] and electron microscopy.[14]

It is clear that the broad deployment of both human-based and beyond-human workflows for automated experiments (AE) necessitates development of a methodology to monitor the progression of the active learning process interacting with experimental physical systems, and tune the AE progression on a time scale affordable for humans. This includes both developing tools for explainability of the AE in real time and post experiment, and the creation of frameworks that allows human intervention on the time scale and decision-making levels amenable to human scientists.

Here we introduce the concept of the after-experiment AE forensics and human-in-the-loop interventions. We demonstrate this approach for scanning probe microscopy (SPM), but this concept is equally applicable for scanning transmission electron microscopy (STEM), materials synthesis in automated labs, and theory exploration over large chemical spaces.

We consider the general process of forensic analysis for AE and introduce (or adapt) key concepts necessary in this case. The central element of experimental active learning is the ML agent interacting with the experimental system in an iterative fashion, both performing the experimentation and updating the state of the ML agent. Initially, the ML agent's state is defined by priors and inferential biases (e.g. a hypothesis list, invariances, pre-trained networks) formed based on human input. Throughout the experiment, the state of the agent is updated in response to the incoming information from the active data generation process (i.e. microscope). Based on the current state and prior information, the agent makes decisions that are communicated to the microscope. This iterative cycle continues until the experimental budget is exhausted, or the predefined goal is achieved. At the end of the experiment, the change in the state of the agent represents the knowledge gained during the experiment. Correspondingly, we define AE forensics as the analysis of the decision-making at each step of the experiment, comparing the decisions made by the agent in the actual knowledge state compared to the decision that would have been made by the fully trained agent. It is also important to note that the forensic assessment can include a human component, i.e., based on the experimental results at the final (for forensic) or intermediate (for human in the loop intervention) the human operator can choose to change the nature of the information available to the agent or policies that guide the decision-making and explore the experimental path from this perspective.

Here, we discuss the forensics for a specific case of a deep kernel learning automated experiment workflow for SPM. Traditionally, the SPM is fully operated by human operators with a certain small number of stages amenable to automation. The typical SPM imaging workflow starts with the sample selection, sample preparation and loading, and microscope tuning (often automated). With the sample loaded and microscope initiated, a human operator initiates overview and spectroscopy scans, making subsequent decisions based on result at each step. As a consequence of numerous operations performed by human operators manually, the experiment is a laborious and time-intensive process. Furthermore, the decision-making process by human operators is often slow and is heavily biased by human decision making. For the latter, human operator makes decision based on specific hypothesis and goals (introducing bias). Similarly, it is challenging for human beings to understand high-dimensional datasets and their relationship in a short time frame.

In contrast, ML algorithms can analyze high-dimensional datasets quickly, e.g., deep learning can learn a relationship between thousands of structural image patches and spectroscopic properties. Bayesian methods,[25,26] as exemplified by deep kernel learning,[16,24,27,28] allow an active learning process, making decisions based on past acquired information. When implementing a workflow with deep kernel learning in an operating SPM, the microscope is able to perform the measurement, process data, make decisions to move probe, initiate scan or spectra, automatically without human intervention. This largely surpasses the speed of measurements carried out by human operators, accelerating physics discovery. However, tuning this process requires understanding the decisions made by the ML agent, and adjusting the policies that guide these decisions. Here we discuss the deep kernel learning analysis of a pre-acquired dataset with known ground truth and the forensics. We have chosen the analysis of a pre-acquired data set since only in this case different experimental paths can be compared; however, the methodologies developed here can be implemented on active microscope in a straightforward fashion. We also note that the same logic can be applied to other experiments as well, e.g., the molecular design, processing trajectory optimization, etc.

We have chosen band excitation piezoresponse spectroscopy (BEPS) data of a $PbTiO_3$ (PTO) thin film as a model ground truth data. We have chosen the analysis as a pre-acquired data set since only in this case different experimental paths can be compared; however, the methodologies developed here can be implemented on active microscope in a straightforward fashion. The PTO thin film were grown on a (001) $KTaO_3$ substrate with a $SrRuO_3$ conducting layer. The band excitation piezoresponse force microscopy[29] (BEPFM) imaging of this film are shown in Figure 1 (a-c). The black domains in Figure 1 (a) amplitude image are *a*-domains with in-plane polarization and the bright domains are *c*-domains with out-of-plane polarization, indicating the presence of a typical *a-c* domain structures in this PTO film. The dark and bright domains in Figure 2b phase image show the antiparallel $c^+$ and $c^-$ out-of-plane polarized domains. In addition, resonance frequency image in Figure 1 (c) also shows the ferroelastic a-c domains. In our earlier work, we revealed a mutual interaction among the image channels in BEPFM via causal physical mechanism analysis.[28] We also implemented deep kernel learning in operating SPM for autonomous experiment to explore the structure-property relationship in this sample,[16,24] e.g., the relationship between ferroelectric domain structure and polarization-voltage hysteresis loop. The latter are traditionally implemented as a grid spectroscopy measurement,[30] which provides us low resolution image showing domain structure and corresponding spectroscopy of polarization-voltage hysteresis at each pixel.

Here, we use a BEPS data acquired in our previous work[24] as a model to illustrate DKL analyses and establish forensics workflow. The BEPS data is shown in Figure 1 (d-f). Figure 1 (d) is the image showing ferroelectric domain structure, Figure 1 (e) shows two example patches that will be used as structural data in DKL analyses, and Figure 1 (f) shows two example polarization-voltage hysteresis corresponding to these patches.

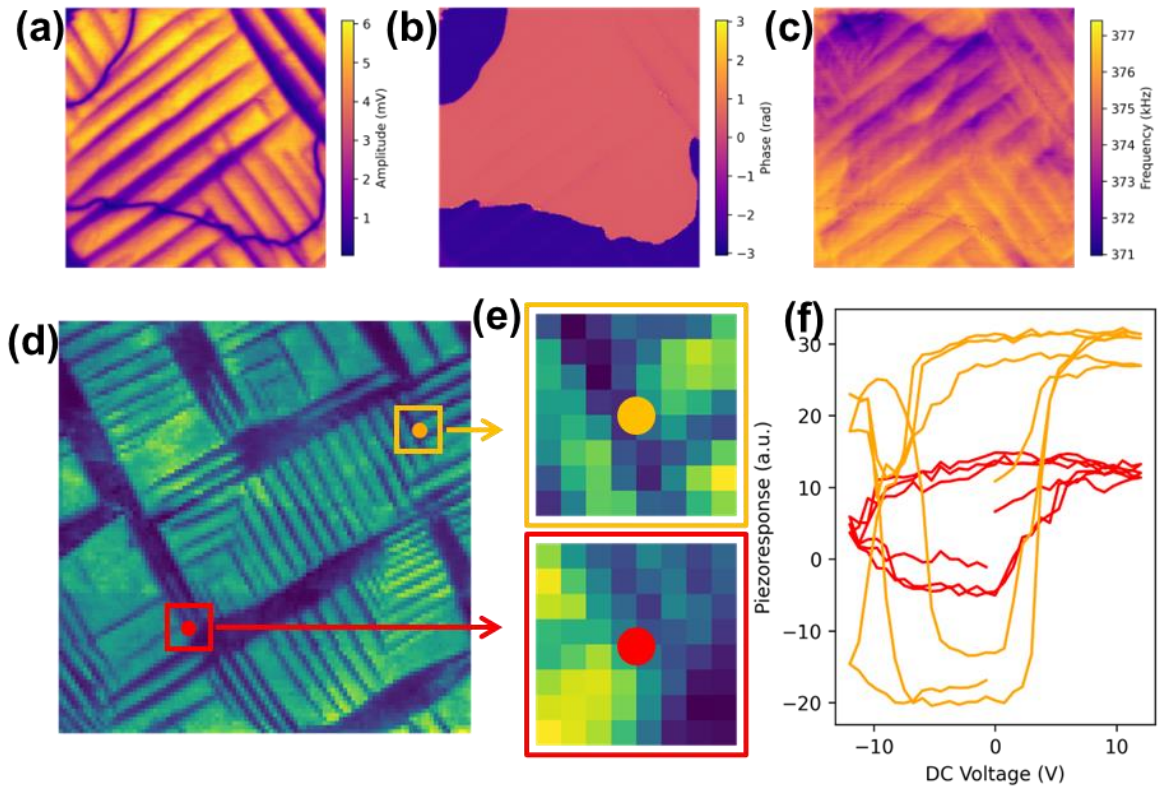

**Figure 1.** Band Excitation Piezoresponse Force Microscopy (BEPFM) and Spectroscopy (BEPS) results of model sample $PbTiO_3$. (a-c), BEPFM amplitude, phase, and frequency images showing ferroelectric and ferroelastic domain structures in PTO thin film. (d-f) BEPS results of PTO thin film, (e-f) shows two example domain structures and corresponding spectroscopic behaviors. Panel (d-f) is reproduced with permission from Ref[24].

In the DKL experiment shown in Figure 2, the agent has access to the global. The global image can be sampled at the individual locations $[x_i, y_i]$ to return the local image patch (i.e., known structure) and allows for local measurement (i.e., measured spectrum) at the same location. The (image patch, spectrum) pair represents the feature and target of the DKL process, where at the initial stage all the features and small number of targets are available. During the active learning process, the algorithm identifies the next feature (i.e., location $[x_{i+1}, y_{i+1}]$) to perform the measurement at, provides the coordinates to the microscope, receives new data and appends it to the target list, and updates the model with the expanded list of targets. The general task of the exploratory DKL algorithm is to learn the relationship between feature and targets by iteratively selecting the next sampling location, and attempt to arrive at this in the smallest number steps. For a ferroelectric material, this corresponds to learning the relationship between the local domain structure (patch) and local hysteresis loo, but this approach applies to any measurement pairs such as images and spectra in STEM-EELS, optical images and indentation curves, or even molecular structure and functional properties in chemical spaces.

In order to arrive at the structure-property relationships in the smallest number of steps, the DKL method is used as a basis for a Bayesian Optimization[31] process. In the DKL BO framework, the chosen characteristic of the spectrum (functional) or evolution of spectrum sequence (mutual entropy, etc.) defines the reward function for a DKL active learning process. The goal in AE is then to maximize this function. For example, DKL AE can be used to discover the microstructural elements that correspond to the largest area under on-field or off-field hysteresis loops in PFM,[24] regions with highest intensity of edge plasmons in the STEM EELS measurements,[14] or highest internal electric field in the 4D STEM.[32] It is also important to note that DKL AE is an example of beyond-human workflows, since the correlative relationship between the spectra and structure is learned in real time from sequential data acquisition process.

The balance between the exploration and exploitation on the BO framework is set via the choice of the acquisition function built upon the scalarizer function. Here we have implemented and investigated three acquisition functions in this work, namely expected improvement (EI), upper confidence bound (UCB), and prediction maximum uncertainty (MU). Specifically for DKL-PFM experiment, the DKL is trained by image patches $IP_i$ and measurements $M_i$ (chosen as scalarizer function applied to spectrum) at locations $[x_i, y_i]$; the trained DKL makes prediction of unmeasured locations with known structural image patches. The scalarizer function can be e.g., the area under the hysteresis loop, built-in bias (i.e., offset of the loop in the voltage axis), or any other characteristics of the spectrum. Then, the acquisition function derives the next measurement location $[x_{i+1}, y_{i+1}]$ based on DKL prediction, the next measurement location is chosen as an argmax of the acquisition function. We define the sequence of patches, locations, and measured spectra as the **experimental trace**. It is important to note that the decision making in the DKL AE is based on a single scalar characteristics of the measured spectra (i.e., scalarizer) rather than the full spectrum. However, availability of full spectrum allows us to incorporate counterfactual analyses as will be illustrated below.

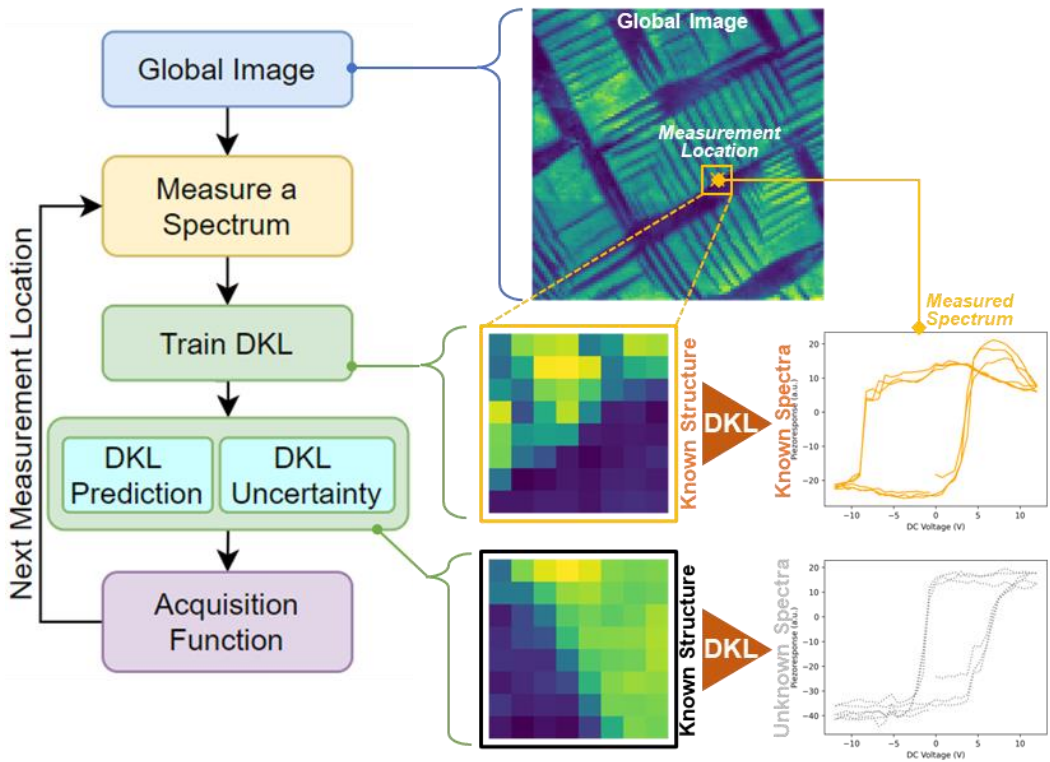

**Figure 2.** Deep Kernel Learning (DKL) workflow. DKL predicts the values of scalarizer function (i.e., characteristic of spectrum that is of interest to experimentalist, such as area under hysteresis loop, magnitude of signal at certain voltage, etc.).

Here we introduce the forensic analysis framework for DKL experiment, comprising the (a) regret analysis including the acquisition function component analysis and counterfactual decision making, (b) trajectory analysis and feature discovery, and (c) global latent trajectory analysis.

In regret analysis, we compare the knowledge gain along actual discovery path followed during the AE and the gain that would have been achieved with fully trained model if it were available in the beginning of the experiment. To illustrate this concept, here we have defined three DKL models as shown in Figure 3. The **live DKL model** is the active model during DKL experiment, which is trained and updated at each exploration step. This model has learned only from the experimental data available from the beginning of the experiment to the current step. The **final DKL model** is the model trained on all data sampled during DKL experiment, e.g., in a 200 step DKL experiment, the final DKL model is trained on 200 sampled image patches and corresponding spectroscopic properties. The **full DKL model** is trained by all data in the model BEPS data. Note that the full DKL model can be trained only when the ground truth data is available, whereas active learning experiment allows access to live and (in the end) trained DKL models.

During the forensic regret analysis, the predictions of the final DKL and complete DKL models are compared to the learning process of the live DKL model. The regret is defined as the difference between predicted scalarizer of live DKL model and final DKL model:

$$Regret_i = Prediction_i^{live\ DKL} - Prediction_i^{final\ DKL}$$

The regret evolution for three acquisition functions in Figure 3b. Here, the solid line in Figure 3b indicates the mean $Regret_i$, and the shadow indicates the deviation of $Regret_i$ across the data set (i.e. the standard deviation of the DKL uncertainty image).

The learning process of live DKL models with different acquisition functions are shown in Figure 3c-e. The DKL prediction uncertainty is used as metrics of learning here. The uncertainty of final DKL model and complete DKL model are showed as a comparison. Here, the prediction uncertainty characterizes how well can the model predict the scalarizer value from the observe domain pattern. The black line illustrates the evolution of the uncertainty during AE, whereas the shaded region is the distribution of uncertainties within the image. For the UCB and EI acquisition functions the model learn faster. However, for EI there is a clear inflection point suggesting discovery of the new type of behaviors. Experimentally, the smoother learning process can be achieved by increasing the exploratory component of the chosen acquisition function. Curiously, the variability of the predicted behaviors is small in the beginning of the learning process, grows at the intermediate steps when the model discovers new classes of behaviors, and starts to decrease for the second half of the training. The regret function (comparison of the final and live model) for this data set shows high noise level and allows to identify the stage of maximal learning. Finally, the mean prediction component can further be evaluated to characterize the behavior of the optimization part of the algorithm (not shown).

We note that the **learning curves** in (c-e) can be evaluated during the experiment and hence represent the indicators based on which hyperparameter tuning (choice of acquisition function, tuning the exploration and exploitation within acquisition function, addition of random exploration e.g. via epsilon-greedy policies) can be introduced.

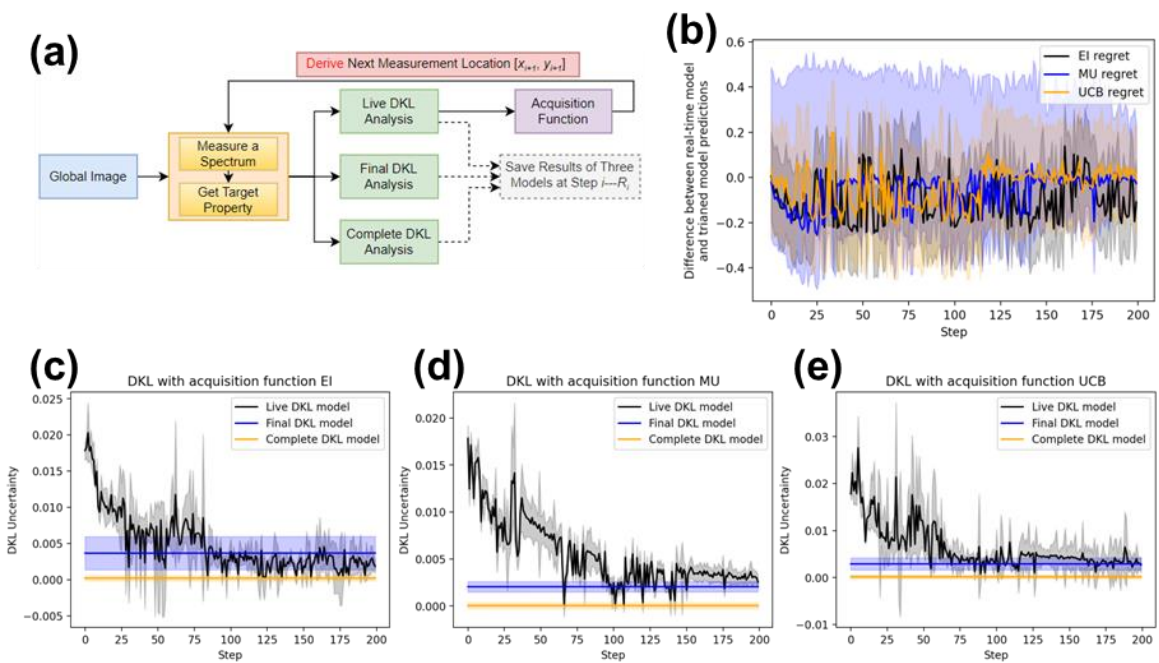

**Figure 3.** Regret analysis of DKL AE. (a) the workflow of regret analysis, where the final DKL and complete DKL models are reference models. The final DKL model is the DKL model after 200 steps of exploration, and the complete DKL model is trained with all available data. (b), Regret of DKL with different acquisition functions. The regret is defined as $Rg_i = P_i^{real-time\ DKL} - P_i^{trained\ DKL}$, where the solid line is the mean of $Rg_i$ and the distribution is the deviation of $Rg_i$ at step i. (c-d) comparison of live, final, and complete DKL prediction uncertainty as a function of step, (c-d) shows results of DKL with three different acquisition function respectively; here the solid line show the average uncertainty and the shadow shows the deviation of uncertainty.

The deeper insight into decision making process can be made based on the analysis of the components of the acquisition function, including predicted value and uncertainty. Therefore, the DKL predicted value and uncertainty of next measurement point are shown in Figure 4 as a function of step. The ground truth of next measurement point is also shown in Figure 4 as a comparison. The observed traces show very high noise levels common for active learning tasks. Note that this high noise level is present both in the prediction and the ground truth data. However, the background trend of the decreasing the predictive value (i.e., BO minimizes the scalarizer function) is clearly seen and the rate of learning for different acquisition functions can be clearly deduced as the bottom envelope of observed behaviors.

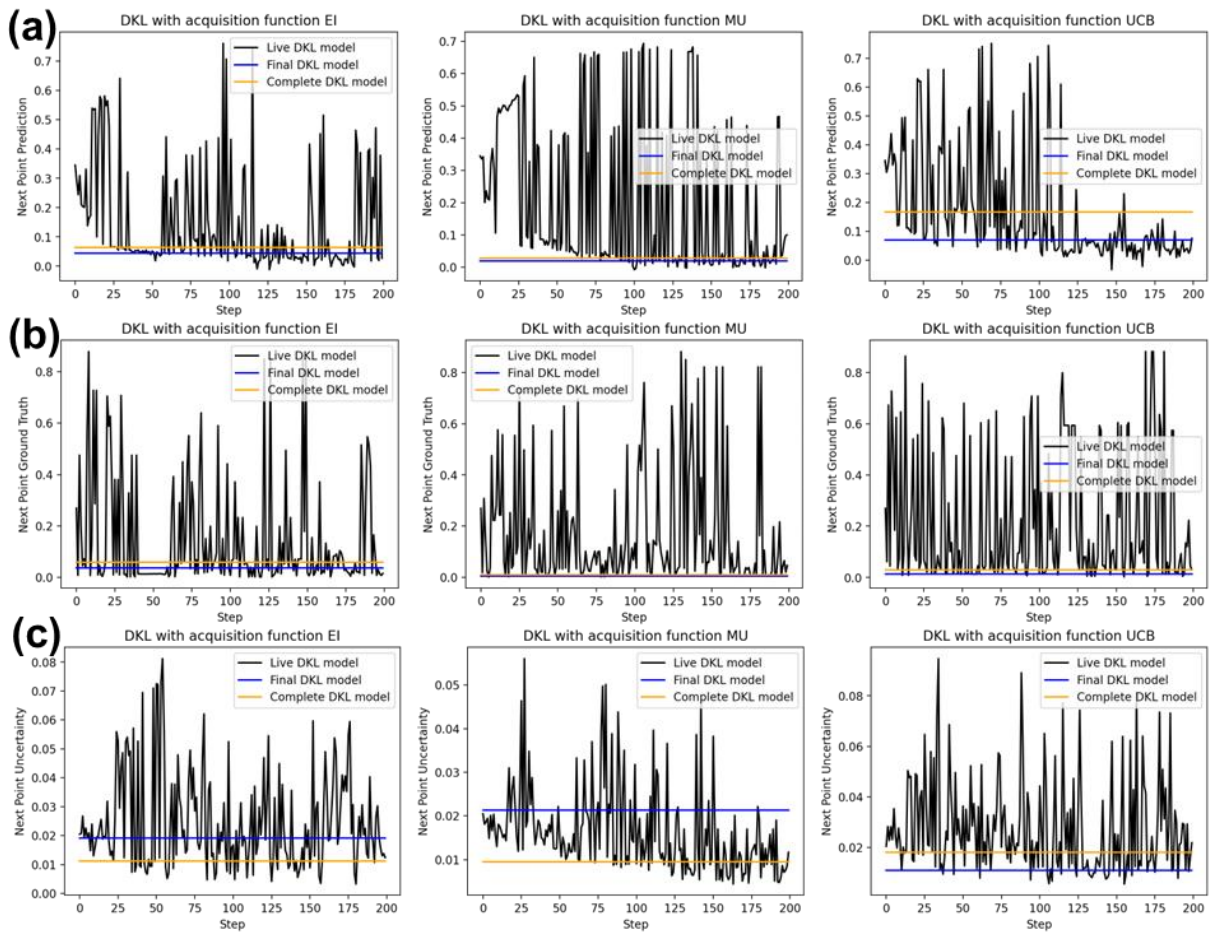

**Figure 4.** Comparison of live, final, and complete DKL prediction and uncertainty for next measurement location. (a) shows DKL prediction of the value of next measurement location and (b) shows the ground truth of predicted next point. (c) shows the uncertainty of next point.

The second component of the forensic analysis that we introduce is the counterfactual forensic analysis. We recall that the progression of the deep kernel active learning (DKAL) is driven by the chosen characteristic of the spectra defined via the scalarizer function. For example, the scalarizer can be the work of switching, nucleation bias, or imprint for hysteresis loops in PFM, or any other functional of the spectrum defined via analytical expression or neural network. The prediction and uncertainty of the scalarizer form the acquisition function. DKL experiment uses a selected scalarizer to guide the exploration.

Here, we introduce the counterfactual analysis, defined as how the action at each step change would if the scalarizer had been chosen to be different. Obviously, this analysis is possible only at the step-by-step basis, since for subsequent steps the selected image patch would have been different. The reason the counterfactual analysis is possible at each step is because the actual experiment collects the full spectrum data, and hence any scalarizer function for actual step can be evaluated.

This counterfactual analysis allows to determine whether the sampled points based on target property 1 are helpful for exploring target property 2. To illustrate this concept, we have performed DKL exploration with loop area from polarization-voltage hysteresis, as shown in Figure 5 (b). This creates **experimental trace**, meaning the sequence of the sampling points, corresponding image patches, and hysteresis loops. With the experimental trace for loop area as a scalarizer created, we have loaded the sampling points and used loop height or coercive field to perform DKL analysis, as shown in Figure 5 (c). In addition to DKL counterfactual analysis that uses DKL sampling points based on a difference target property, we also performed DKL analysis using random sampling points as a comparison. DKL counterfactual and random analyses results are shown in Figure 5 (d).

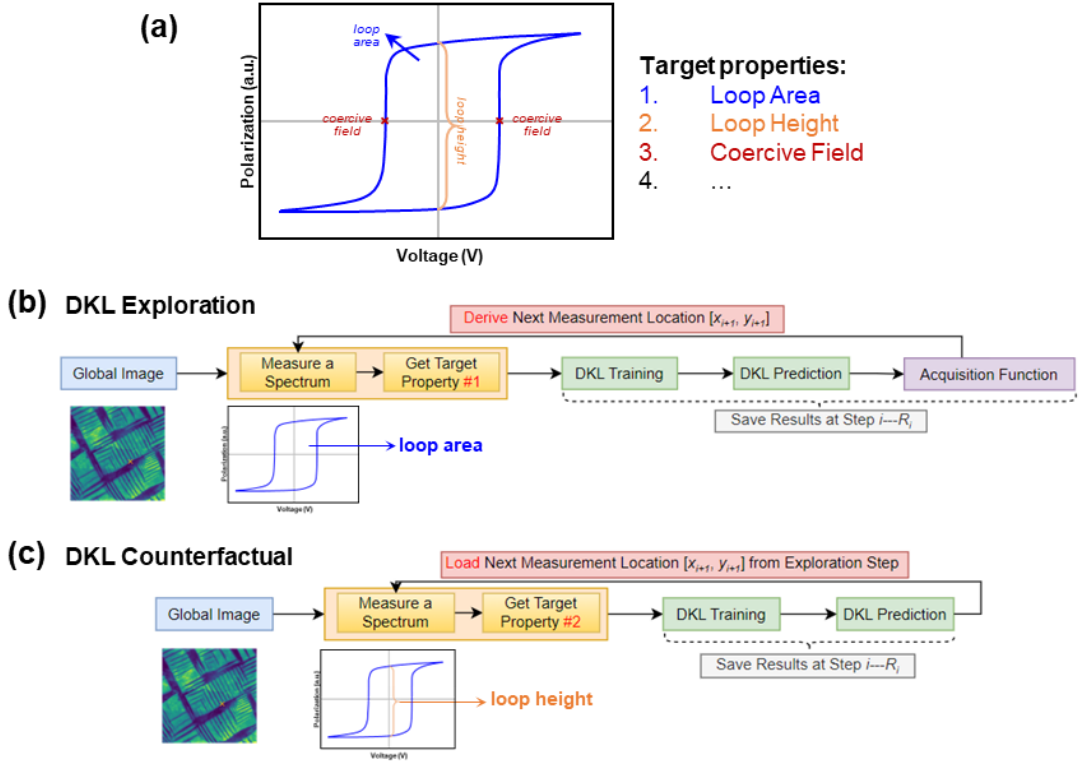

**Figure 5.** DKL counterfactual analysis. (a-c) workflow of counterfactual analysis. (a) multiple target properties encoded in a spectroscopic data. (b) DKL exploration based on one target property determine the measurement location at each step. (c) in counterfactual analysis, another target property will be used to perform DKL analysis and the measurement location at each step is loaded from the corresponding exploration step. (d) DKL counterfactual analysis results, where the DKL exploration is guided by target property of loop area, and counterfactual analysis target properties are loop height and coercive field. DKL analysis based on random location is also plotted as a comparison.

As a measure of the counterfactual experiment progression, we compare the spatial distribution of the predicted images with the ground truth. Here, we generate the DKL prediction image of target functionality, and calculate the structural similarity index (SSID) between the DKL prediction and ground truth. We expect that if the physical behaviors described by different scalarizers are strongly correlated the SSID values will be high, whereas if the differ the SSID will be low.

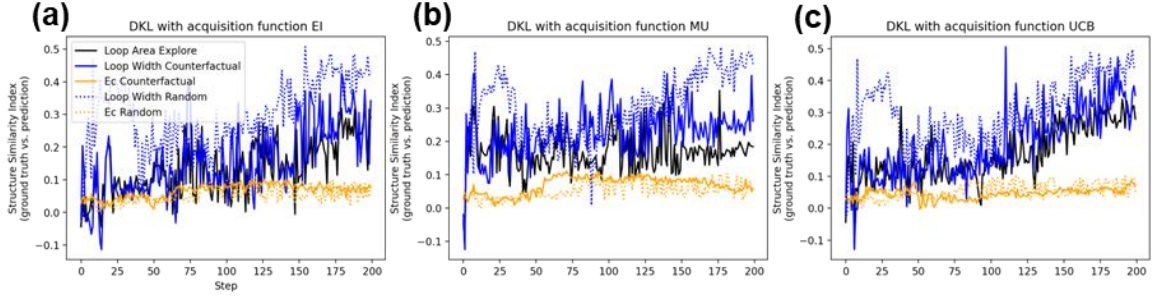

**Figure 6.** DKL counterfactual analysis results, where the DKL exploration is guided by target property of loop area, and counterfactual analysis target properties are loop height and coercive field. DKL analysis based on random location is also plotted as a comparison. (a-c) show counterfactual analysis results of DKL with different acquisition function. The DKL counterfactual analysis process is shown in Figure 5.

Shown in Figure 6 are the DKL counterfactual analysis results with different acquisition functions. The experimental trace is created by performing DKL exploration with loop area from polarization-voltage hysteresis as the target property. Then, we loaded the experimental trace and used loop height and coercive field to perform counterfactual analysis. The SSID evolution of DKL exploration (with loop area as target property) and DKL counterfactual with loop height as target property are similar, as shown in Figure 6. However, the evolution of DKL counterfactual with coercive field as target property is slightly different from the DKL exploration. This is most likely because that the loop area and loop height originate from similar physical mechanisms, i.e., remnant polarization magnitude. In contrast, coercive field and loop area are tied to different physical mechanisms.

We further proceed to define **trajectory analysis** and **feature discovery**. We note that AE in physical imaging traces a certain trajectory in the image plane of the system. Given that the global image is available before the DKL experiment, this trajectory can be visualized and examined in real space both in real time and after the experiment.

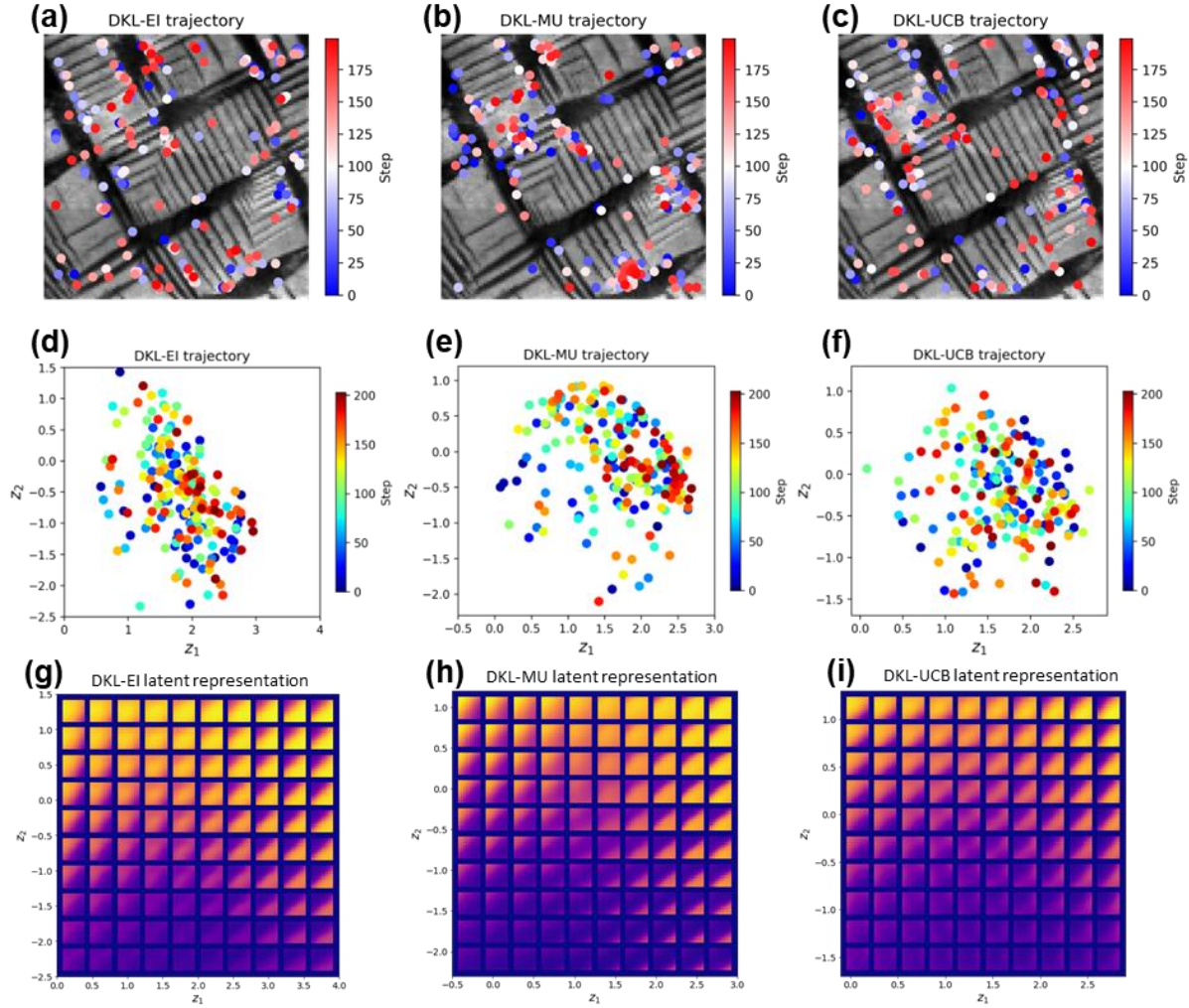

**Figure 7.** Trajectories of DKL exploration with different acquisition functions and evolution of latent component $Z_1$, $Z_2$ as a function of step. (a-c) Trajectory of DKL exploration, where the exploration locations are shown on the structural image, the spots color represents step. (d-f) Trajectory in latent space. (g-i) rVAE latent representations.

Shown in Figure 7 (a-c) are the trajectories of the DKL exploration. The exploration locations are plotted in the structural image in sequence from blue to red. The examination of the real-space trajectory yields a powerful real-time and forensic tool to monitor the progression of the AE. For example, concentration of the experimental points in a certain part of image plane to full exclusion of other regions often suggests the effects of instrumental cross-talk (e.g. tilt). Secondly, visual examination of the trajectory vs. the structural image allows direct identification of the microstructural elements that carry functionalities of interest discovered by the DKL. For example, in Figure 7 (a) and (c) it is clear that many of the experimental points have been chosen at the ferroelastic domain boundaries between large *a* and *c* domains. At the same time, the sampling of the small *a-c* domain stripes is very sparse, suggesting that functionality of interest (i.e. hysteresis loop area) does not manifest strongly in these regions even though they comprise most of the sample surface. Note that this analysis can be further extended towards human-in-the loop analysis, where supervised ML can be used to identify objects of interest (human provided goal and labels) and next round of the AE will be focused only on these specific features. This approach has been demonstrated for the analysis of grain boundaries in hybrid perovskites.[33]

During the automated experiment the ML agent learns which image patches correspond to optimization of the scalarizer function. To explore the dynamics of this learning process and explore what are the discovered features, we introduce the latent analysis with rotationally invariant variational autoencoders (rVAE) on the full experimental trace.[34-36] The rVAE disentangles the factors of variation in all image patches into latent variables, in this case, the latent space shows the structural variations that somehow relevant to physical features (e.g., domains). Shown in Figure 7 (d-f) are the trajectories of DKL exploration in rVAE latent space. The corresponding latent representations are shown in Figure 7 (g-i) and allow to identify the physical features related to latent variables. Note that the analysis above is based on the data contained in the experimental trace and becomes available after the experiment.

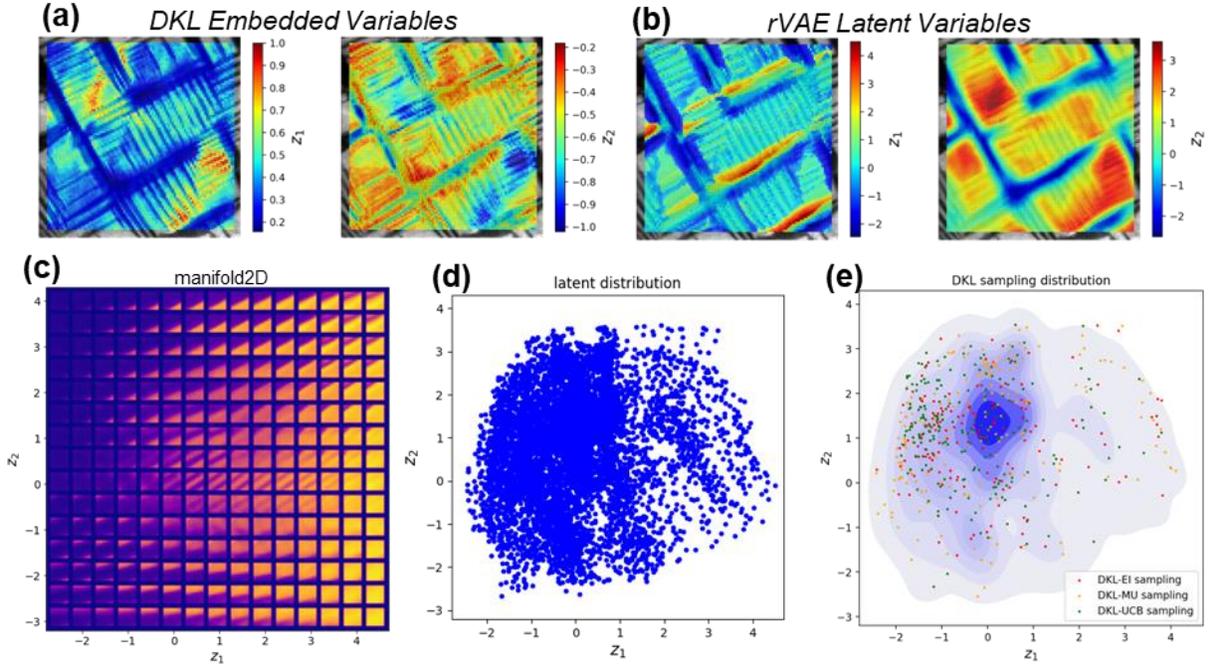

**Figure 8**. rVAE analysis of DKL sampling. (a) DKL embedded variables (b) rVAE latent variables. (c) rVAE analysis of all image patches and the DKL sampling distribution. Here, (d) is latent distribution of the system, i.e., all image patches represented in the latent space. (e) is the superposition of the sampled patches on the kernel density estimate of the full latent distribution of the system, showing which of the regions were sampled for different acquisition functions. These data can be color-coded by the number of point (not shown).

Finally, to explore the discovery process we introduce **latent trajectory analysis**. In this case, we utilize the fact that global image and hence full collection of the image patches is available before the experiment. This allows the latent space of the system to be constructed via the suitable (invariant) VAE. The image patches that become available as the experimental trace can be visualized in the corresponding latent space, whereas the trajectory in the global latent space and evolution of latent variables along the experimental path can be visualized in real time. Note that these latent variables are different from the DKL latents, since the DKL embeddings are updated at each experimental step (and their meaning can be determined only via backpropagation through the DCNN)

The latent trajectory analysis is shown in Figure 8a-b. Figure 8a shows the correlation of DKL embedded variables and structural factors, and Figure 8b shows the correlation of rVAE latent variables and structural factors, where there is some similarity between DKL and rVAE variables. Figure 8c-d indicates the distribution of DKL samplings in the rVAE latent space. Note that for all three acquisition functions, the AE samples the regions in the latent space corresponding to the *a-c* domain walls, whereas the central peak of the kernel density estimate corresponding to most usual microstructures (dense ferroelastic domain patterns) remain unaddressed. This approach illustrates which microstrucutres give rise to the thought behaviors. Note that additional insight into this discovery process can be derived by plotting the time dependence of the latent variables, time-coding trajectories for a single acquisition function in the latent space, etc.

Here we want to note that the structural factors can be modified by changing the image patch size, correspondingly the DKL exploration trajectory and samplings will also be changed. We believe this change is tied to the physics involved in the image patches with different size. Note that the discovery process can be further explored by exploring time dependence of the latent codes corresponding to the patches along the experimental trace.

**Table 1.** Summary of the definitions for forensic descriptors.

| Characteristic | Definition | Availability |
|---|---|---|
| Global image | Initial structural data set available before DKL experiment. Used to create patches for DKL training | Before |
| DKL latents | The latent variables encoding the structural information in the patches | During** |
| Scalarizer function | Function defining what characteristic of spectrum guides Bayesian Optimization | Before* |
| Acquisition function | Function combining DKL prediction and uncertainty of the scalarizer function | Before* |
| Policy | Principle for selection of next path. Simplest policy is maximization of acquisition function, but can be more complex including epsilon-greedy or switch between multiple scalarizers or acquisition functions. Human in the loop intervention tunes some aspect of the policy | Before* |
| Experimental trace | Collection of patches (and their coordinates) and spectra derived during experiment. Trace and global image are the results of AE SPM. | During |
| Live DKL model | DKL model in the state corresponding to the *n*-th experimental step | During |
| Final DKL model | DKL model in the state corresponding to the end of the experiment | After |
| Complete DKL model | DKL model trained on the full data set (if available from grid measurements, etc). | |

| Regret analysis | The difference between predictions of live DKL model and final DKL model after the whole experiment (i.e., after 200 steps in this work) | During** and After |
|---|---|---|
| Learning curve | Change of the DKL uncertainty (mean and deviation), indicative of the predictability of the patch-scalarizer relationship | During |
| Counterfactual scalarizer | The availability of full spectral data as a part of experimental trace allows to estimate what the BO step would be if scalarizer were chosen to be different | During |
| Trajectory analysis | Real-time trajectory of the probe that can be represented in the global image plane | During |
| Feature discovery | Analysis of the latent variables and latent representations of image patches and spectra in the trace. Here, we realize only patch analysis but extension to spectra is straightforward. | After |
| Latent trajectory analysis | Analysis of the experimental trajectory in the latent space of the full collection of the image patches derived from the global image | During** |
| | | |

* Denotes parameters that provide controls for human in the loop intervention
** Denotes observables that can be naturally monitored during the DKL experiment to make human-in-the-loop decisions. Note that strictly speaking all after-experiment descriptors can be evaluated on the fly, but represent more difficult to interpret and intervene upon behaviors.

The forensic analysis represented above illustrates the collection of descriptors available during and after the DKL automated experiment that provide insight into the progression of the training (predictive uncertainties), rate of the Bayesian Optimization of target functionalities, and real- and latent space discovery trajectories. Given the rich nature of the information contained in the experimental trace, these methodologies can be developed further using multiple tools developed in the context of static ML and Bayesian optimization. For example, the analysis can be extended to exploring the emergence of correlations between image and spectral data in trace via linear (canonical correlation analysis)[37] and VAE-based method. The patches in trace or spectra can be used as labels for the semi-supervised analysis of the global structural data. Multiple opportunities further emerge for the AE policies, including the introduction of multi-objective optimization for multiple scalarizers, changing policies during the experiment, etc. We defer the analysis of these opportunities for further experimental effort.

Finally, we consider the opportunities for the human-in-the loop interventions in the automated experiment. Table 1 summarizes the definitions for the real-time and forensic descriptors in the AE. The indicators such as learning and regret curves and real-space and latent trajectories can be visualized in real time during the experiment, and provide strong signals on the progress in predictability and discovery during AE and the nature of uncovered structural elements and functional behaviors. At the same time, AE allows easy access to the control parameters. The BO pathways can be tuned via the selection of the scalarizer function that can be chosen from the pre-populated list or dynamically tuned during the experiment (e.g. signal averaged over selected spectral band). The balance between exploration and exploitation can be tuned via the acquisition function, again selecting from the list or tuning the weight coefficients in UCB, etc. Similarly, the random exploration can be added via epsilon-greedy term.

Several of the parameters (e.g. related to final DKL model or feature discovery) are available at the end of the DKL experiment. Rigorously, these can be updated throughout the experiment. However, we believe that their interpretability makes them idea for the human in the loop interventions, where the experiment is paused and human operator tunes the experimental policies. We also note that this analysis can be further extended to introduce additional knowledge during the experiment, for example use deep convolutional network to perform the image segmentation and run DKL on segmented (rather than raw) data. We hope that the notebooks provided with the manuscript will allow the broad experimental community to explore these opportunities.

To summarize, here we proposed an implemented the forensic analysis for the automated experiment in Scanning Probe Microscopy. Here, this analysis is implemented on full pre-acquired data set, thus allowing comparison of the dissimilar experimental policies. However, it can be implemented on any microscope equipped with previously reported DKL workflows. Overall, this analysis allows monitoring the progression of the exploratory and exploitative descriptors during the AE, and introduces the strategies for human-in-the-loop intervention based on the target and policy controls.

The proposed approach can be extended to all other imaging-spectroscopic methods, including Scanning Transmission Electron Microscopy – Electron Energy Loss Spectroscopy (STEM-EELS), STEM – 4D STEM, optical microscopy or scanning electron microscopy combined with nanoindentation, and multiple chemical imaging methods. The only requirement for the applicability of forensic analysis in its present form is the availability of the structure-spectra pairs as the basis for the imaging process. From instrumental side, the common denominator is the control hyper-language that gives access to the probe position and enables initiation of imaging and spectroscopic scans.

We further note that this methodology can be applied for more complex scenarios, including exploration of the parameter space of the theoretical models or composition and processing spaces in automated materials synthesis, as well as chemical spaces for organic molecules or biomolecules. However, in these cases the measures for exploration and representation will depend on the structure and correlations in the corresponding parameter spaces, necessitating the development of domain-specific descriptors. Overall, we believe that proposed framework opens the pathway both to interpretable automated experiment, AE monitoring, and human-in-the-loop interventions.

**Acknowledgements**

This effort (SPM measurement, forensic workflow development, data analysis) was primarily supported by the center for 3D Ferroelectric Microelectronics (3DFeM), an Energy Frontier Research Center funded by the U.S. Department of Energy (DOE), Office of Science, Basic Energy Sciences under Award Number DE-SC0021118. SPM experiments were done at the Center for Nanophase Materials Sciences (CNMS), which is a US Department of Energy, Office of Science User Facility at Oak Ridge National Laboratory. S.V.K. acknowledges support from the Center for Nanophase Materials Sciences (CNMS) user facility which is a U.S. Department of Energy Office of Science User Facility, user project no. CNMS2022-B-01642.

**Conflict of Interest Statement**

The authors declare no conflict of interest.

**Authors Contribution**

S.V.K., R.K.V, M.A.Z, and Y.L. conceived the project. Y.L. developed forensics workflow based on DKL and rVAE from M.A.Z. All authors contributed to discussions and the final manuscript.

**Data Availability Statement**

The analysis codes that support the findings of this study are available at https://github.com/yongtaoliu/Forensics-DKL-BEPS.

The DKL codes are implemented using AtomAI package at

https://github.com/pycroscopy/atomai.

The rVAE codes are implemented using pyroVED package at

https://github.com/ziatdinovmax/pyroVED